\documentclass{article}

\usepackage{PRIMEarxiv}

\usepackage[utf8]{inputenc} % allow utf-8 input
\usepackage[T1]{fontenc}    % use 8-bit T1 fonts
\usepackage{hyperref}       % hyperlinks
\usepackage{url}            % simple URL typesetting
\usepackage{booktabs}       % professional-quality tables
\usepackage{amsfonts}       % blackboard math symbols
\usepackage{nicefrac}       % compact symbols for 1/2, etc.
\usepackage{microtype}      % microtypography
\usepackage{lipsum}
\usepackage{fancyhdr}       % header
\usepackage{graphicx}       % graphics
\usepackage{subfig}         % subfigures
\usepackage{amsmath}       % 添加这一行来支持\text命令

\title{AgentSociety Challenge: Designing LLM Agents for User Modeling and Recommendation on Web Platforms}

\author{
  Yuwei Yan$^{1,2\dagger}$\\
  \And  
  Yu Shang$^{1\dagger}$\\
  \And
  Qingbin Zeng$^{1}$\\
  \And
  Yu Li$^{1}$\\
  \And
  Keyu Zhao$^{1}$\\
  \And
  Zhiheng Zheng$^{1}$\\
  \And
  Xuefei Ning$^{1}$\\
  \And
  Tianji Wu$^{3}$\\
  \And
  Shengen Yan$^{3}$\\
  \And
  Yu Wang$^{1}$\\
  \And
  Fengli Xu$^{1,*}$\\
  \And
  Yong Li$^{1,*}$
}

\begin{document}
\maketitle

\begin{abstract}
The AgentSociety Challenge is the first competition in the Web Conference that aims to explore the potential of Large Language Model (LLM) agents in modeling user behavior and enhancing recommender systems on web platforms. The Challenge consists of two tracks: the \textit{User Modeling Track} and the \textit{Recommendation Track}. Participants are tasked to utilize a combined dataset from Yelp, Amazon, and Goodreads, along with an interactive environment simulator, to develop innovative LLM agents. 
The Challenge has attracted 295 teams across the globe and received over 1,400 submissions in total over the course of 37 official competition days. 
The participants have achieved 21.9\% and 20.3\% performance improvement for Track 1 and Track 2 in the Development Phase, and 9.1\% and 15.9\% in the Final Phase, representing a significant accomplishment.
This paper discusses the detailed designs of the Challenge, analyzes the outcomes, and highlights the most successful LLM agent designs. 
To support further research and development, we have open-sourced the benchmark environment at \url{https://tsinghua-fib-lab.github.io/AgentSocietyChallenge}.
\end{abstract}

\keywords{LLM Agent, Recommender System, User Modeling, Web Platforms}
\footnote{\\
$^1$ Tsinghua University\\
$^2$ The Hong Kong University of Science and Technology (Guangzhou)\\
$^3$ InfinigenceAI\\
$^\dagger$ Both authors contributed equally to this work. \\
$^*$ Corresponding authors. E-mail: fenglixu@tsinghua.edu.cn,liyong07@tsinghua.edu.cn.
}

\section{Description}
With the rapid advancement of web technologies, human social interactions are increasingly intertwined between the physical world and cyberspace. The web has evolved from a mere information exchange platform to a dynamic medium for modeling and influencing human behaviors. Web platforms, in particular, provide rich data capturing how users access information, make decisions, and interact with content and other users. These data offer invaluable insights into human dynamics, especially in understanding user intent and optimizing information retrieval (IR) systems for personalized online service.
In parallel, large language models (LLMs) have demonstrated exceptional capabilities in reasoning and prediction tasks~\cite{wei2022chain,yao2024tree,chen2024large}, which are now widely applied in user behavior modeling and recommendation systems~\cite{wu2024survey,liao2024llara}. LLM-based agents~\cite{shang2024agentsquare,shao2025division} further extend these capabilities by effectively simulating complex, generative human behaviors~\cite{park2023generative,gao2024large,zhang2024generative,ding2024understanding}. 
This advancement has profound implications for behavior modeling, paving a new way to model user preferences, engagement patterns, and decision-making processes. 

LLM agents possess unique strengths that extend beyond traditional deep learning models. One key advantage is their capability for commonsense reasoning~\cite{wei2022chain,yao2024tree,shang2024defint,xu2025towards}, allowing them to make few-shot predictions based on minimal data and effectively addressing data-sparse tasks such as recommendations for users with limited historical records.
Additionally, their proficiency in zero-shot role-playing~\cite{park2023generative,zhang2024agentcf} enables them to simulate user behaviors in diverse contexts and adapt to new scenarios with in-context information. This makes them particularly valuable for tasks such as simulating context-sensitive and personalized user intent and behaviors. However, there's still a critical challenge in bridging these technological advancements with practical, useful tools for information retrieval and recommendation systems.
The \textbf{AgentSociety Challenge} is proposed to address this gap by asking participants to leverage the simulation and decision-making power of LLM agents to tackle two fundamental IR challenges: simulating user behaviors and developing personalized recommendation systems. The challenge consists of two parallel tracks: 

\textbf{User Modeling Track.}
In this track, participants are tasked to design agents to simulate user reviews and star ratings, focusing on simulating user behavior when facing specific items by leveraging their historical actions and accessible environmental data. 

\textbf{Recommendation Track.}
In this track, participants will develop LLM agents that act as a recommendation assistant~\cite{zhang2024agentcf,zhang2024generative}. The agent should provide tailored recommendations for users based on historical interactions and retrieved information. 

AgentSociety Challenge aligns closely with the core themes of the Web Conference by addressing fundamental problems in web technologies and user-centered Computing. The competition draws the participation of 295 teams and receives over 1,400 submissions over the course of 37 official competition days. Rooted in areas like information retrieval, recommendation systems, and user modeling, this challenge leverages cutting-edge LLM agents to simulate and predict complex human behaviors in dynamic web-enabled environments. 

\section{Dataset and Evaluation}
\subsection{Dataset}
The proposed challenge utilizes three open-source datasets, including Yelp, Amazon, and Goodreads, to evaluate agents. These datasets were carefully chosen for their relevance and diversity in capturing real-world user interactions, making them ideal for building LLM-based agents for user modeling and recommendation.

\begin{itemize}
    \item \textbf{Yelp Dataset}: This dataset~\cite{asghar2016yelp} includes millions of user reviews, business information, and user feedback such as ratings and engagement metrics. 

    \item \textbf{Amazon Dataset}: This dataset~\cite{hou2024amazon} contains millions of reviews for a wide range of products, offering insight into user preferences in the e-commerce context. 

    \item \textbf{Goodreads Dataset}: This dataset~\cite{wan2018goodreads} includes user ratings, reviews, and metadata for books, capturing user preferences and interactions in the book community. 
\end{itemize}

\subsection{Environment Simulator}
One key capability of a Web Agent is its ability to acquire information and data from interactive web platforms. To adequately assess LLM agents, it is essential to furnish an environment that reflects these intricacies. This environment empowers agents to perform tasks such as user modeling and personalized recommendations within an interactive and realistic setting.

\begin{figure}[t!]
  \centering
  \includegraphics[width=0.8\columnwidth]{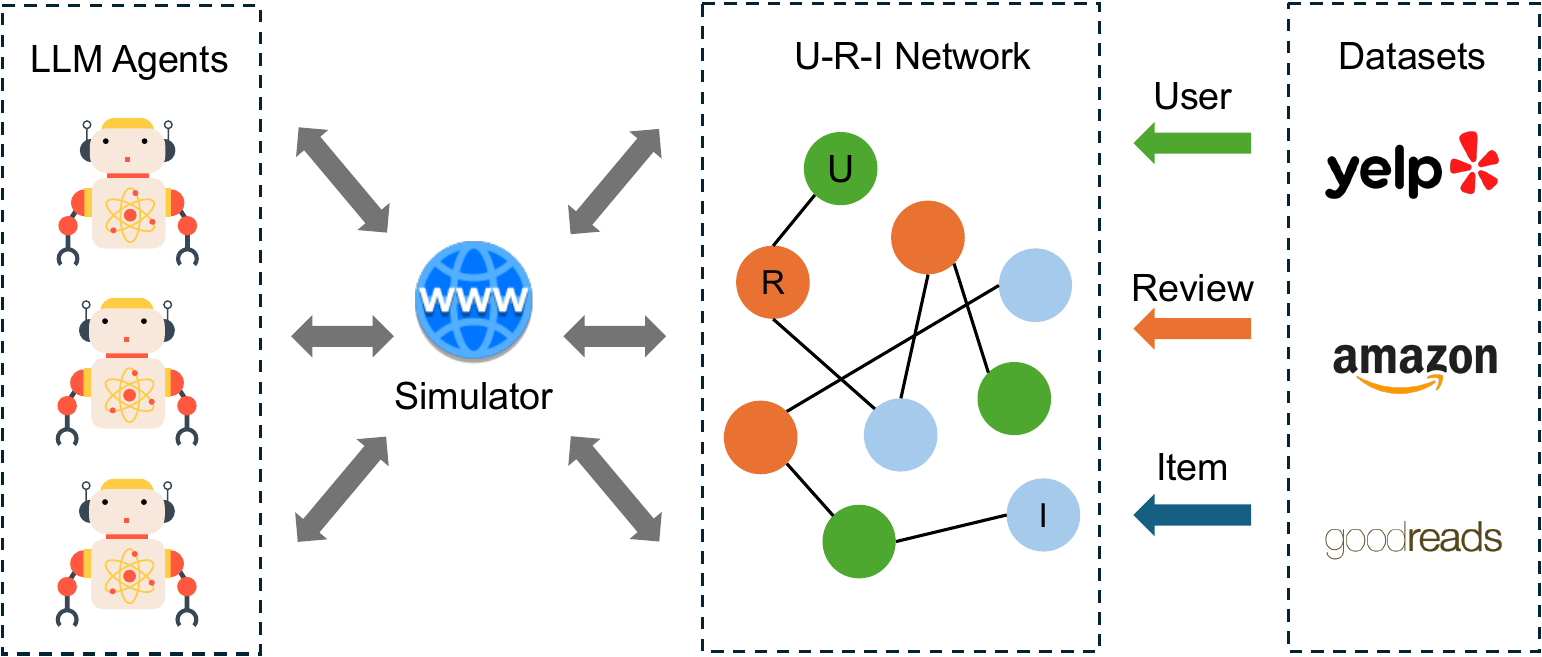}
  \caption{Framework of environment simulator.}
  \label{fig:simulator}
\end{figure}

To that end, we have built a simulator that controls all retrieval actions from LLM agents. The core of the simulator used in the challenge is the \texttt{InteractionTool}. This tool constructs an interactive environment comprising a network of users, reviews, and items, allowing agents to access historical data as needed. By leveraging these datasets, agents can simulate user behaviors, including generating reviews and ratings and produce personalized recommendations based on contextual information. This configuration enables a thorough evaluation of agent performance in tasks that closely resemble real-world applications. The framework of the simulator is shown as Fig.\ref{fig:simulator}.

\subsection{Evaluation}
\textbf{User Modeling Track.}
In this Track, the performance of participants' agents will be evaluated through quantitative metrics, focusing on the accuracy of user behavior predictions. 

Preference Estimation:
The preference estimation accuracy is evaluated through the mean absolute error (MAE) of the predicted star ratings compared to the actual ratings. The MAE is defined as:
\[
MAE = \frac{1}{N} \sum_{i=1}^{N} |{\hat{s}_{ni} - s_{ni}}|
\]
where \(N\) is the total number of reviews, \(\hat{s}_{ni}\) is the predicted star rating, and \(s_{ni}\) is the ground truth star rating\cite{wang2024recmind}. A lower MAE indicates a better performance in estimating user preferences.

Review Generation:
The review generation accuracy is measured through a combined metric that considers Emotional Tone Errors, Sentiment Attitudes Errors, and Topic relevance errors. The overall review generation error is given by: $\text{Error} = 1 - (0.25 \times \text{Emotional Tone Error} + 0.25 \times \text{Sentiment Attitude Error} + 0.5 \times \text{Topic Relevance Error})$
where each individual error is calculated as follows:

\begin{itemize}
  \item Emotional Tone Error. A predefined emotion classifier model\cite{user@emotion} and the error is the MAE of the normalized emotion scores between the predicted and actual emotions. 
  \item Sentiment Attitude Error. The sentiment attitude of the review text is analyzed using the MAE of sentiment scores derived from sentiment analysis tools.
  \item Topic Relevance Error. A predefined embedding model\cite{user@sentence} and the error is measured using the cosine similarity between the predicted and actual text embeddings.
\end{itemize}

Overall Quality:
The overall quality of the reviews is evaluated by averaging the Preference Estimation and Review Generation errors.

\textbf{Recommendation Track.}
The Recommendation Track focuses on developing LLM agents that generate personalized recommendations for users. The performance of the recommendation agents is evaluated through the precision of the ranking, with a focus on the top N hit rate. 

Ranking Accuracy:
Ranking accuracy is evaluated using the top N hit rate, where \(N = 1, 3, 5\). This metric measures how often the ground truth item appears in the top N of the ranked list of 20 candidate items. The formula for the Top N Hit Rate is:
\[
HR@N = \frac{1}{T} \sum_{t=1}^{T} \mathbb{I}(p_t \in P_t^{\hat{}}(N))
\]
where \(T\) is the total number of test tasks, \(p_t\) is the ground truth item, and \(P_t^{\hat{}}(N)\) is the set of top N recommended items for task \(t\). The indicator function \(\mathbb{I}(p_t \in P_t^{\hat{}}(N))\) equals 1 if the ground truth item is in the top N recommendations and 0 otherwise.

\textbf{Two Phase Evaluation.}
The evaluation procedure is divided into two phases: Development Phase and Final Phase (Top 20 teams per track). In the Development Phase, simulated groundtruth is utilized to assess participants' agents, while a combination of simulated groundtruth (40\%) and real groundtruth (60\%) is employed in the Final Phase. 
This design aims to prevent overfitting by ensuring that agents are not solely optimized for public available data. Simulated groundtruth represents new or unseen users (data), offering a broader scope for testing how agents handle variations that might not be captured in real-world data alone. 
A blocking mechanism is implemented to conceal the real groundtruth according to the specific task requirements, which ensures that agents infer and generate outputs without direct access to the groundtruth. 
The simulated groundtruth generation agents are available at \url{https://github.com/tsinghua-fib-lab/AgentSocietyChallenge/tree/main/GTsimulation}.

\section{Competition Outcome Analysis}
\subsection{Submission Performance Statistics}
\begin{figure}[t!]
  \centering
  \subfloat[Development Phase.]{\includegraphics[width=0.5\linewidth]{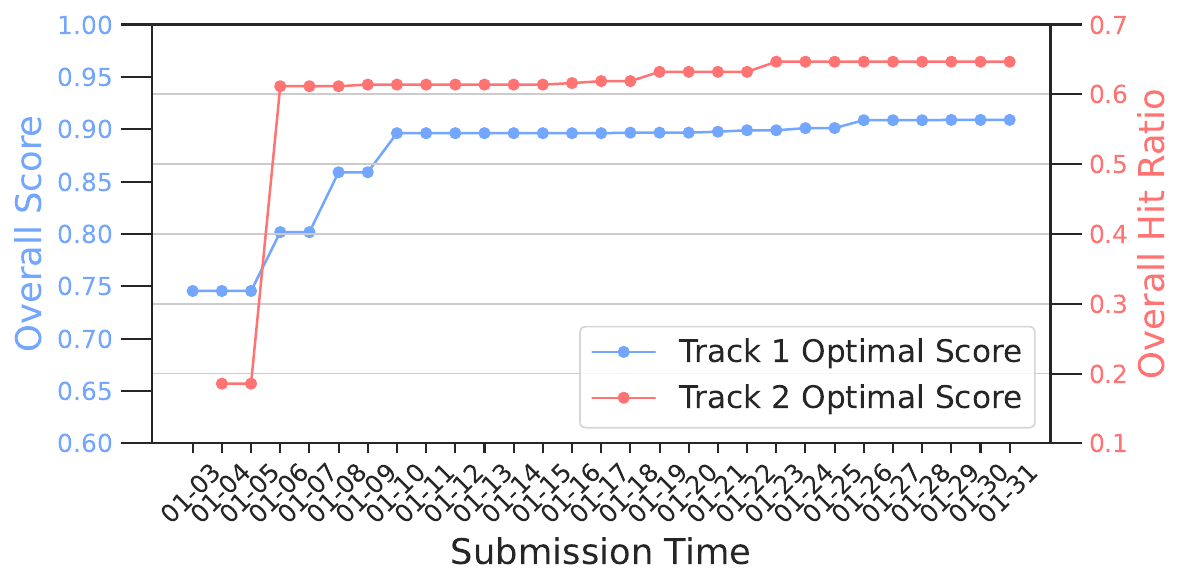}}
  \subfloat[Final Phase.]{\includegraphics[width=0.5\linewidth]{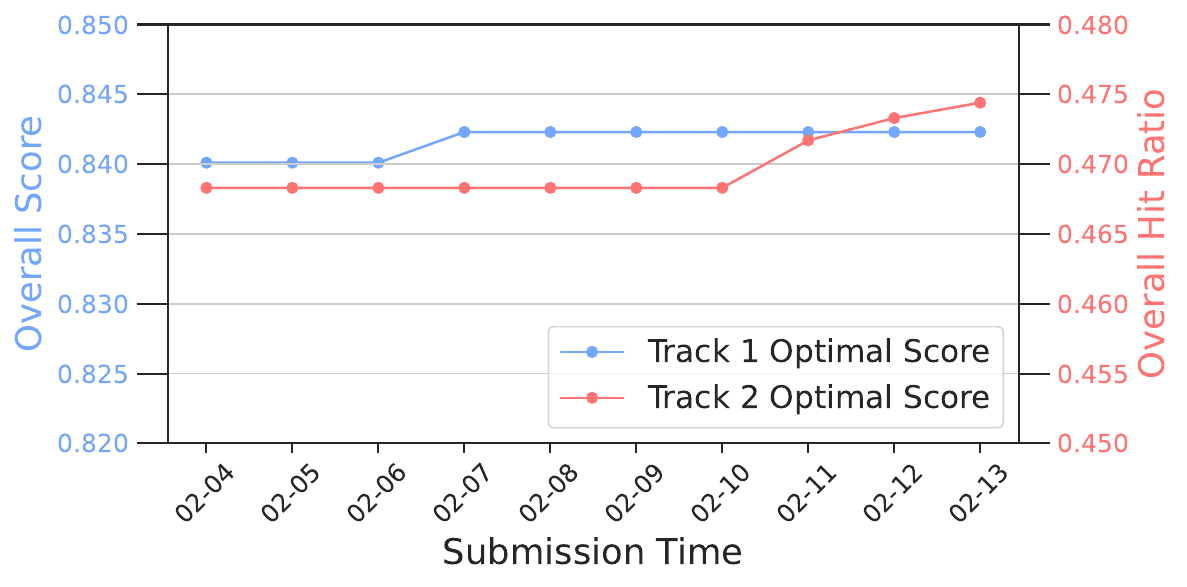}}
  \caption{Performance improvement in Development Phase (a) and Final Phase (b).}
  \label{fig:dev_final_st}
  \end{figure}

There are 295 teams registered for the challenge. 
During the Development Phase, Track 1 received a total of 503 submissions, while Track 2 received 405 submissions. 
After the Development Phase, the top 20 teams from each track advanced to the Final Phase. 
During the Final Phase, Track 1 received 262 submissions, while Track 2 received 282 submissions.

Fig.~\ref{fig:dev_final_st} presents the daily submission performance statistics for both tracks during the development and Final Phases. 
Overall, the performance of the submitted agents gradually improved over time. 
At the beginning of the challenge, we provide an official agent for reference on both tracks.
Early in the Development Phase, participants' designed agents are primarily around the baseline. 
However, as the challenge progressed, several designs emerged that significantly surpassed the performance of the baseline agent.
Moreover, we have found that the designed agent-based recommendation methods generally outperform traditional deep-learning recommendation models (e.g., NCF~\cite{he2017neural}), highlighting their significant potential.

\subsection{Textual Environment Simulator}
Here we provide an analysis of the simulated groundtruth used for evaluation. First, we aim to ensure that the simulation groundtruth can objectively assess agents' performance, thereby reducing the risk of overfitting to specific real-world test conditions. Second, we validate that the simulated groundtruth accurately reflects actual user behaviors and preferences, demonstrating its potential to enhance traditional methods, such as deep learning-based models.
\begin{figure}[t!]
\centering
\subfloat[Track 1.]{\includegraphics[width=0.4\linewidth]{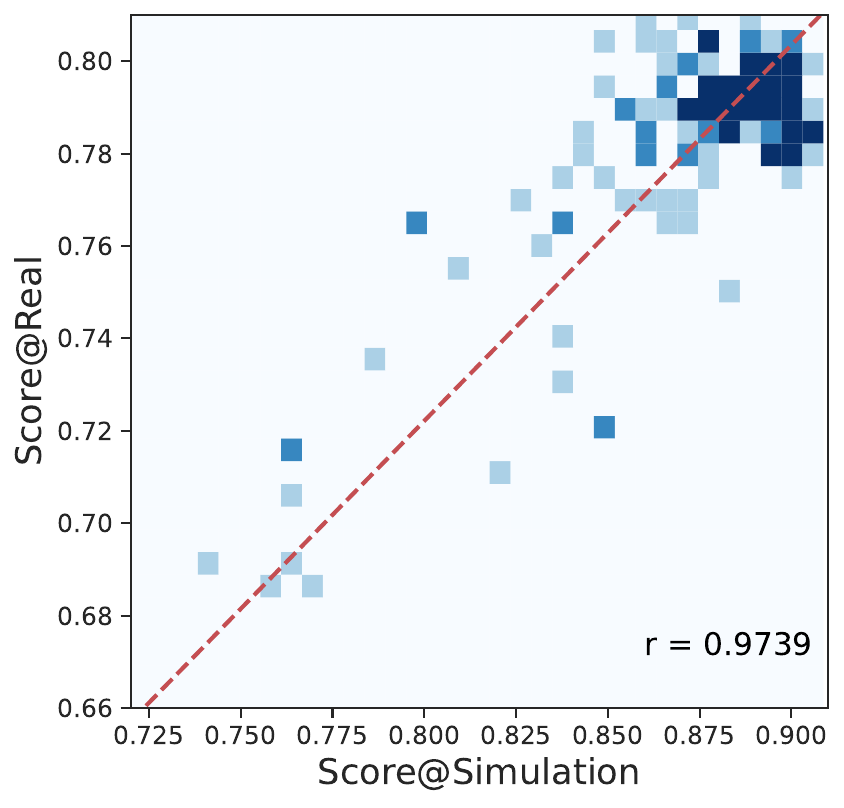}
}
\subfloat[Track 2.]{\includegraphics[width=0.4\linewidth]{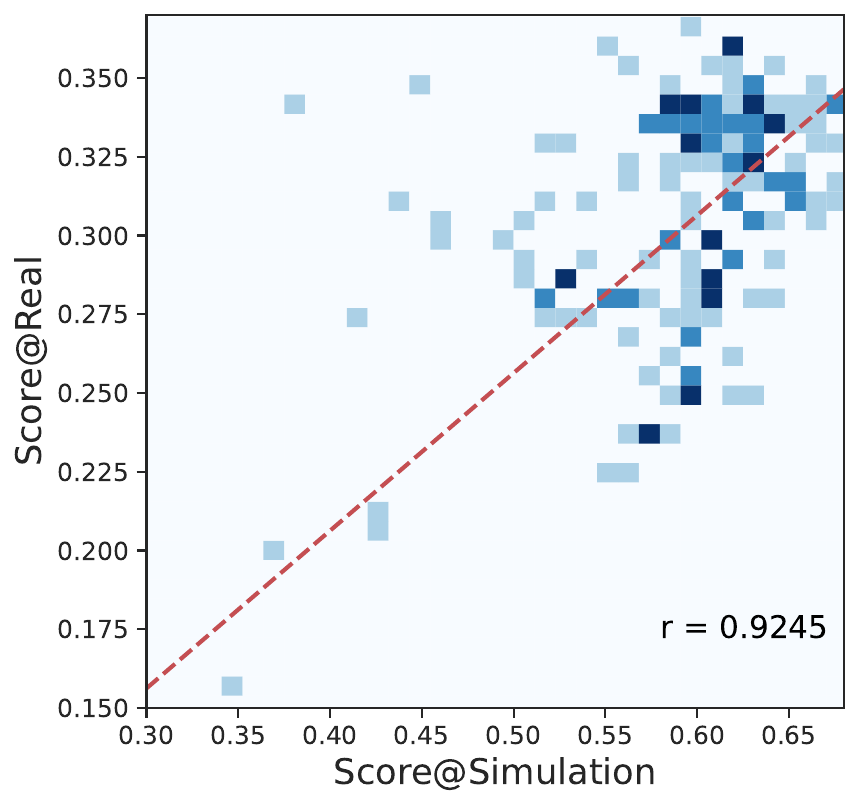}}
\caption{Correlation between agent performance on simulated and real groundtruth on Track 1 (a) and Track 2 (b).}
\label{fig:corr_sim_real}
\end{figure}

\textbf{Correlation between simulated and real groundtruth for evaluation.}
We analyze the correlation between the performance of submitted agents (179 agents on Track 1 and 196 agents on Track 2) during the Final Phase of our generated simulated and real test groundtruth. 
The results are shown in Fig.~\ref{fig:corr_sim_real}, from which it can be observed that there's a strong correlation between the performance on simulated and real groundtruth, with a Pearson correlation coefficient of 0.9739 for Track 1 and 0.9245 for Track 2. 
This suggests that the generated simulated groundtruth is reliable and meaningful, thus the agent performance on the simulated groundtruth can provide a strong prediction of their real-world performance.

\begin{figure}[t!]
\centering
\subfloat[Mixed data v.s. Real data B.]{\includegraphics[width=0.4\linewidth]{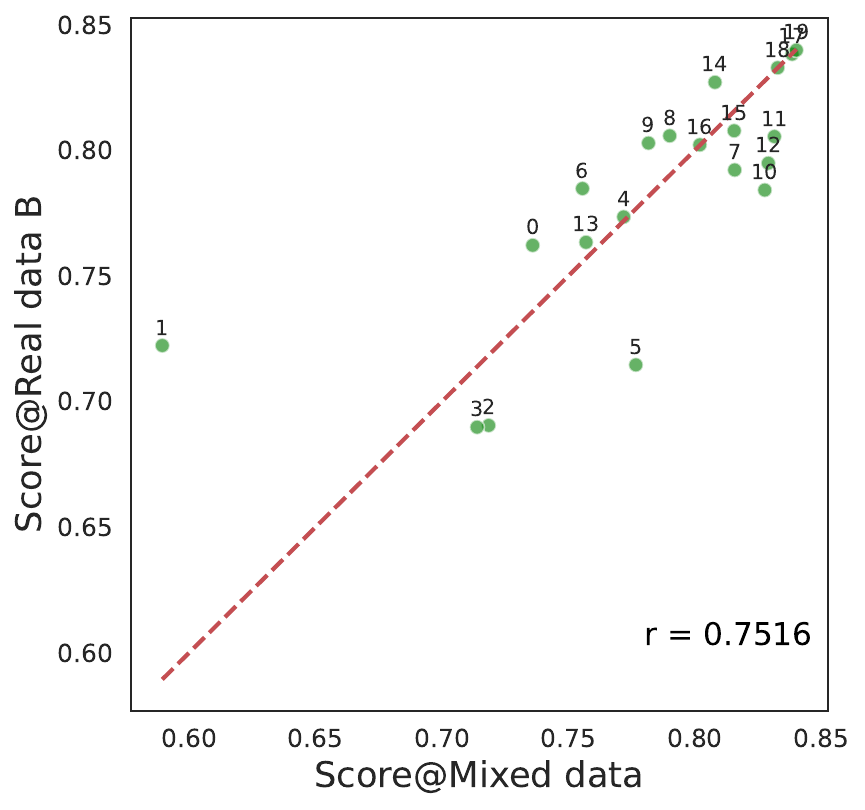}}
\subfloat[Real data A v.s. Real data B.]{\includegraphics[width=0.4\linewidth]{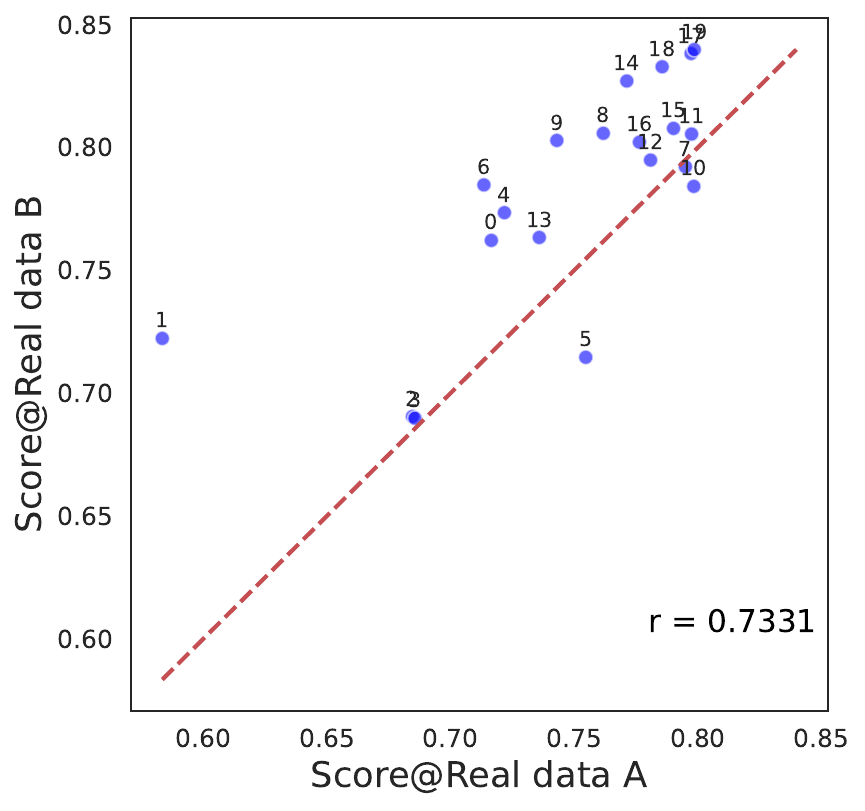}}
\caption{Generalization ability comparison on Track 1.}
\label{fig:corr_track1}
\end{figure}

\begin{figure}[t!]
\centering
\subfloat[Mixed data v.s. Real data B.]{\includegraphics[width=0.4\linewidth]{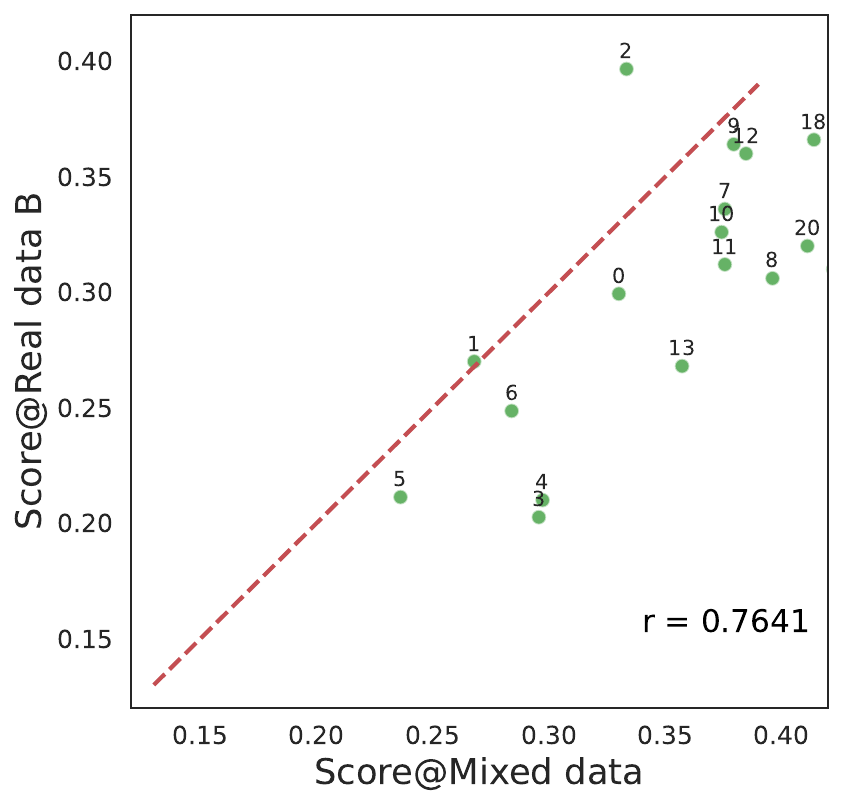}}
\subfloat[Real data A v.s. Real data B.]{\includegraphics[width=0.4\linewidth]{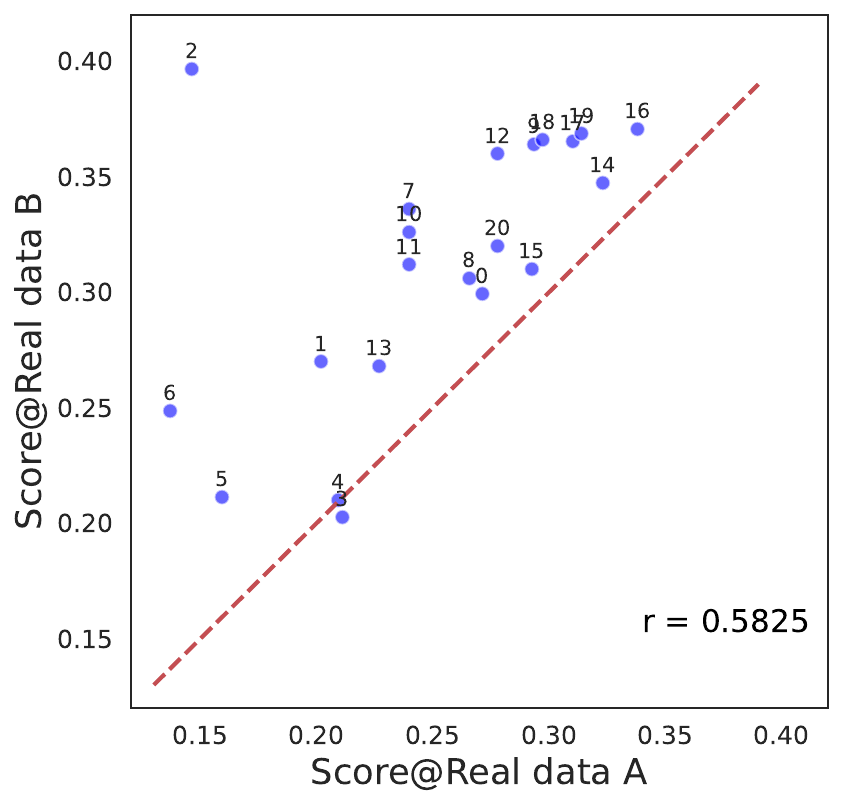}}
\caption{Generalization ability comparison on Track 2.}
\label{fig:corr_track2}
\end{figure}

\textbf{Generalization ability of mixed and real groundtruth.} 
We evaluate the agents' performance on the following three types of datasets: (1) a mixture of simulated and real groundtruth (referred to as "Mixed Data"), (2) a sample of real groundtruth (referred to as "Real Data A"), and (3) another sample of real groundtruth (referred to as "Real Data B").
We then compare the agent performance correlation of Mixed Data and Real Data A against Real Data B to assess their generalization capability. 
The results on Track 1 and 2 are shown in Fig.~\ref{fig:corr_track1} and \ref{fig:corr_track2}, respectively.
For Track 1, the Pearson correlation coefficient between Mixed Data and Real Data B is 0.7516, while the correlation between Real Data A and Real Data B is 0.7331. 
In Track 2, the correlation between Mixed Data and Real Data B is 0.7641, whereas the correlation between Real Data A and Real Data B is lower, at 0.5825. 
The results suggest that the use of mixed simulation-real groundtruth leads to better performance prediction of the agents' real-world capabilities compared to purely using real groundtruth.
For instance, Agent 2 on Track 2 performs poorly on Real Data A but excels on Real Data B. 
Such a discrepancy can be effectively mitigated when using mixed groundtruth for evaluation.
This finding suggests that incorporating simulated groungtruth helps provide a more robust evaluation of the agents' real performance.

\begin{figure}[t!]
  \centering
  \includegraphics[width=0.7\columnwidth]{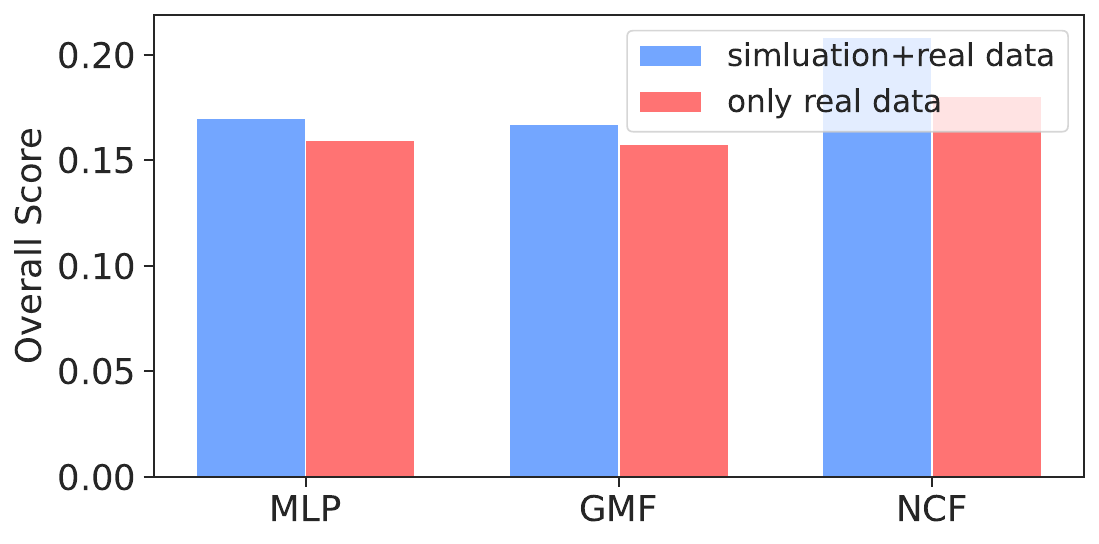}
  \caption{Performance comparison of deep-learning-based recommendation methods on pure real groundtruth and mixed groundtruth (with simulated groundtruth).}
    \label{fig:ncf}
\end{figure}

\textbf{Effect of simulated groundtruth on classic deep learning models.} 
We tested three deep learning-based recommendation models from ~\cite{he2017neural}—NCF, GMF, and MLP on both simulated and real groundtruth. 
The results from Fig.~\ref{fig:ncf} show that incorporating a certain amount of simulated groundturth into the training process improved the model performance compared to training solely on real groundtruth. 
This indicates that the simulated groundtruth effectively captures user behaviors and preferences, thereby facilitating the learning of user features.

\section{Case Study}
\subsection{Top Agents in User Modeling}
This section analyzes the top three performing agents from the User Modeling Track (named ``ASC'', ``JiuWen'', ``STDYW''). Specifically:

\textbf{Design Pattern.} All agents follow a multi-stage retrieve-plan-generate pipeline. They employ contextual prompt engineering by combining user profiles, item attributes, and historical reviews using platform-specific templates. 

\textbf{Extracted Features.} Each agent employs a unique approach tailored to its objectives, with different strategies for generating ratings, memory handling, and prompt styles. ASC employs a collaborative filtering core strategy, integrating a preference alignment engine and using MDILU with similarity search for memory, along with statistical mean adjustment for rating logic. In contrast, JiuWen uses an aspect-based analysis strategy, with an aspect extractor as its unique module and example-driven memory. Its rating logic is guided by examples, and it employs case-based examples in its prompt style. STDYW, on the other hand, relies on direct LLM generation, with a sanitization module and basic DILU memory approach. Its rating logic is based purely on LLM output, and it uses concise instructions for its prompts.

\textbf{Different Competitive Advantages.} 
The distinctive design choices made by each agent lead to notable competitive advantages. 
ASC's integration of user and item mean ratings with variance is a sophisticated technique for improving rating consistency. By adjusting for the statistical variance of the data, ASC minimizes rating fluctuations, helping to achieve more stable and reliable user behavior predictions. 
JiuWen's design excels in its innovative approach to extracting aspects from business reviews. By breaking down reviews into specific aspects, such as service quality, product features, and customer experience, JiuWen can align user behaviors with fine-grained features. 
Another standout feature is STDYW's combination of advanced user-and-item modeling and knowledge mining. By integrating these components with sophisticated reasoning capabilities, STDYW enhances personalized and context-aware simulations, offering superior predictive accuracy and human-like review generation.

Overall, these implementations highlight that effective user behavior modeling requires balancing LLM capabilities with structured reasoning frameworks. 

\subsection{Top Agents in Recommendation}
In this section, we analyze the designed agents of the top 3 teams on Track 2 (named ``baseline666'', ``RecHackers'', ``DummyAgent''), and summarize their key design insights as follows:

\textbf{Agentic workflow.}
All three agents relied on similar information to guide the ranking process, with slight variations in implementation. The primary elements for ranking include: (1) the user's historical review data, reflecting past preferences; (2) the list of candidate items to be ranked; and (3) detailed item information, helping assess the match between the items and the user's preferences. Specifically: 

\textbf{Item-side feature engineering.}
The first primary innovation in these agents lies in item-side feature engineering. 
Especially, the baseline666 team applied platform-specific feature extraction, ensuring robust and adaptive rankings across different sources. 
For Amazon, they extracted features like item ID, name, stars, review count, and description, among others. For Yelp, the features were more focused, including item ID, name, stars, and review count. For Goodreads, a wider array of features was included, such as authors, publication year, and similar books.

\textbf{Review-side feature engineering.}
Another key point is the feature engineering of review data, involving filtering and selecting the most informative reviews to enhance both user and item descriptions.
For instance, the DummyAgent team adopted platform-specific strategies. For Yelp, they focused on attributes like ``funny'', ``cool'', and ``useful'', along with the review text. For Amazon, they included the publication date and purchase verification. For Goodreads, they extracted additional metadata like review date, the number of votes or comments, and reading status, in addition to the review text and rating.

To summarize, the key design elements can be categorized into three aspects: First, regarding the agent's workflow, a standard and effective approach involves prompting LLMs with user historical reviews, candidate items, item details, and platform-specific information to generate rankings.
Second, extracting platform-specific item attributes plays a critical role in enhancing performance. Lastly, prioritizing the most relevant and informative reviews is crucial for boosting results.

\section{Conclusion}
The AgentSociety Challenge has not only fostered the design of innovative agent-based solutions but also led to significant improvements in simulating user behaviors and enhancing personalized recommendations. Through the challenge, participants introduced diverse strategies that advanced the accuracy of user behavior simulations and personalized recommendation quality. These outcomes highlight the potential of LLM-driven systems in real-world web platforms and set the stage for continued innovation in user-centered computing.

\bibliographystyle{unsrt}   
\bibliography{main}

\end{document}